# MECHANISMS FOR DWARF NOVA OUTBURSTS AND SOFT X-RAY TRANSIENTS

*A critical review* *

J.P. LASOTA

*UPR 176 du CNRS; DARC*
*Observatoire de Paris, Section de Meudon*
*92190 Meudon Cedex, France*

**Abstract.** I review models trying to explain dwarf-nova outbursts and soft X-ray transients. The disc-instability model for dwarf-novae is still in its preliminary state of development: its predictions depend very strongly on the unknown viscosity mechanism. It is also doubtful that a *pure* disc-instability phenomenon will be able to describe *all* types of dwarf-nova outbursts, in particular superoutbursts. The disc-instability model for SXTs suffers from the same difficulties but in addition its predictions are contradicted by observations of transient sources in quiescence. The illuminated mass-transfer model cannot describe correctly the time-scales of SXT events for main-sequence secondaries with masses less than $1 M_\odot$. The existence of at least three systems with $P_{\rm orb} < 10$ hr seems to rule it out as an explanation of the SXT phenomenon.

## 1. Introduction

Dwarf novae and soft X-ray transients (SXTs) are subclasses of respectively cataclysmic variables (CVs) and low-mass X-ray binaries (LMXBs) which undergo, more or less regularly, high amplitude outbursts. CVs and LMXBs are close binary systems in which a Roche-lobe filling low-mass (secondary) star transfers matter to a compact (primary) object: a white dwarf in CVs, a neutron star or a black-hole in LMXBs. The primary accretes matter through an accretion disc, except for strongly magnetized white dwarfs. There is no doubt that oubursts are accretion events and that their site is

---

the accretion disc around the compact objects. The uncertainty concerns the origin and the cause of the outbursts. Are they due to an increased mass-transfer rate from the secondary or to some instability in the disc itself?

In the case of dwarf-nova outbursts the second hypothesis – the disc-instability model – gained popularity after it was realised that accretion disc equilibria become *locally* thermally and viscously unstable due to strong opacity variations caused by partial ionization of hydrogen (see Cannizzo 1993a for a short history of the disc-instability model). In the mass-transfer model the cause of increased mass-transfer rate was rather obscure. One should realise however that the disc instability is local. To obtain a global "limit cycle" behaviour corresponding to observations one has to "tune" the viscosity prescription and/or the vertical disc structure as shown first by Smak (1984). The $\alpha$-disc approximation (Shakura and Sunyaev 1973) fails to give a consistent physical mechanism for dwarf novae outbursts. At present, as I discuss below, it provides mostly phenomenological descriptions of various processes. It is clear that further progress in this domain will require a substantial improvement of the description of viscosity in accretion discs. This is not a very optimistic perspective.

In section 2 I resume the main elements of the disc instability model and some recent work on the subject. Section 3 will be devoted to the origin of so-called superoutbursts in SU UMa-type systems. Particular attention will be given to WZ Sge, a system which seems to be a close cousin of SXTs.

The disc instability model possesses the undeniable merit of offering a description of the main properties of normal (U Gem-type) dwarf novae outbursts. The same model extended to SXTs does not function very well since it is unable, at least in its present version, to reproduce basic observed properties such as recurrence times and accretion rates in quiescence.

In the case of SXTs, illumination of the secondary by X-rays emitted in the vicinity of the compact object provides a mechanism for an increased mass transfer. Unfortunately, time-scales obtained in this type of model cannot apply to SXTs in which the companion star is on the main sequence. The exact cause of SXT phenomena is therefore still uncertain. This problem as well as some recent attempts to combine mass-transfer and disc-instability mechanisms will be reviewed in section 4.

## 2. The disc-instability model

### 2.1. WORKING PRINCIPLES

The local (i.e. at a given radial distance) equilibria of accretion discs are conveniently represented as curves on the accretion rate $\dot{M}$ (or effective temperature $T_{\rm eff}$) surface-density $\Sigma$ plane. For $2500 \lesssim T_{\rm eff} \lesssim 25000$ K the equilibrium curves have a characteristic "$S$" shape. The upper and lower branches of the $S$ correspond to stable, respectively "hot" and "cold" equilibria. The middle branch with a negative slope corresponds to disc equilibria which are thermally and viscously unstable. If the accretion rate at a certain radius is fixed in the unstable regime the system will undergo "limit cycle" oscillations between the lower and the upper branches. The $S$ - shaped curve implies only a *local* instability. To obtain a *global* "limit cycle" the whole disc (in the simplest case) has to behave in a certain way. From the local $T_{\rm eff} - \Sigma$ relation it follows that the surface density in the "cold" (quiescent) state has to be lower than a certain $\Sigma_{\rm max}$, which has the general form

$$\Sigma_{\rm max} = \Sigma_0 r^{b_r} M^{-b_r/3} \alpha^{b_\alpha} \qquad (1)$$

where $M$ is the mass of the central object, $r$ the distance from its center and $\alpha$ the Shakura-Sunyaev viscosity parameter, and $b_r \approx 1$ so that $\Sigma_{\rm max}$ is an increasing function of $r$ whereas in an equilibrium solution $\Sigma$ is a decreasing function of $r$ in the same regime of physical parameters. An accretion disc configuration which is globally on the lower branch will be in a non-equilibrium state when the mass-tranfer corresponds to a unstable equilibrium.

For a constant mass-transfer rate the type of outburst will depend on two characteristic times: the time $t_{\rm accum}$ it takes the matter accumulating at the outer disc to build up a surface density larger than the critical one, compared to the viscous time $t_{\rm vis}$ it takes the matter to diffuse inward and cross the $\Sigma_{\rm max}$ barrier somewhere nearer to the inner disc boundary. If $t_{\rm vis} > t_{\rm accum}$ one obtains a so-called "outside-in" outburst beginning at the outer disc edge; in the opposite case the outburst is of the "inside-out" type (Smak 1984).

When, locally, the disc becomes thermally unstable its system-point on the $\Sigma - T_{\rm eff}$ plane jumps to the upper hot branch. This increases the temperature and viscosity and an abrupt temperature (and density) gradient forms in a local thermal time. This gradient propagates towards the cold disc regions, igniting them on its passage. This gives rise to a global outburst. (In practice, one has to increase $\alpha$ by hand or to use a suitable viscosity prescription.) The density profile in outburst is a decreasing function of $r$, so that outer disc regions are always closer to the $\Sigma_{\rm max}$ barrier. Therefore the cooling wave that brings the disc back to the lower state (below the $\Sigma_{\rm max}$ line) always proceeds from outside (see e.g. Cannizzo 1993a). It is interesting to notice that, according to most model calculations, only a small fraction (10 - 20%) of the matter stored in quiescence actually accretes on to the central body. The main reason for this rather surprising result is the speed at which the cooling front brings the disc back to the cold state, before a significant amount of matter has had time to accrete (Cannizzo 1993a).

## 2.2. THE DISC INSTABILITY MODEL OF DWARF-NOVA OUTBURSTS

As has been shown already in early papers (see Cannizzo 1993a for a short history of the model), time-dependent calculations in the framework of the disc instability model are able to reproduce some properties of dwarf-nova outbursts. The excellent review of the subject by Cannizzo (1993a) ends with a list of eight arguments in favour of the "limit cycle model". Here I would like to stress the weak points of the model. In the next paragraphs I will discuss the claim that *all* categories of dwarf-nova outbursts are due to "pure disc phenomena" in the sense that all outbursts can be modelled assuming a constant mass-transfer rate.

Ludwig, Meyer-Hofmeister and Ritter (1994) have done a systematic study of how the observed features of dwarf-nova outbursts in various systems can be described by the disc instability model. They used a two-alpha viscosity prescription in which viscosity in quiescence and in oubursts is parametrised by two constants $\alpha_{\rm cold}$ and $\alpha_{\rm hot}$ respectively. They concluded that no single viscosity prescription using an $\alpha_{\rm cold}, \alpha_{\rm hot}$ pair can account for all observed properties of dwarf-nova outbursts. In particular it is difficult to get "outside-in" outbursts, whereas in such systems as U Gem and VW Hyi observations show that outburts start in the outer disc region (Smak 1984).

One possible conclusion is that one should modify the viscosity prescription. This can be done either *ad hoc*, by assuming that, for example, $\alpha$ is a function of $r$, or by finding inspiration in the theoretical work on viscosity in accretion discs. The result in both cases is the same: an alpha depending on $r$, $T$, $\dot{M}_T$, the binary system mass-ratio, etc. Among the most popular are: $\alpha = \alpha_0 (H/r)^n$, where $H$ is the disc half-thickness, or just $\alpha \sim r^\epsilon$. The formula $\nu = \nu_0 (r/r_0)^a (\Sigma/\Sigma_0)^b$ was proposed in a study that is supposed to demonstrate the superiority of the disc instability model over the mass-transfer enhancement model (Mineshige, Yamasaki and Ishizaka 1993).

Another difficulty is the so called UV − lag: in some dwarf-novae the rise to outburst of the optical flux precedes that of the UV flux by 0.5 to 1 day. The disc instability model in its standard form is unable to account for this effect because the heating front arrives too fast at the inner, UV-emitting, disc regions (obviously the outburst has to be of the "outside-in" type). This difficulty could be avoided if the inner disc regions were absent during quiescence. The heating front propagating inward would then end its trip at some $r_{\rm in}(quiescence) > r_{\rm in}(outburst)$. The increased viscosity would rebuild the inner UV-emitting region in a viscous time. In this way one obtains the required UV–lag. Two models removing the inner disc have been proposed. Livio and Pringle (1992) show that a weak magnetic field of the white dwarf may have the required effect. Meyer and Meyer-Hofmeister (1994) show that the inner disc may be evaporated through a coronal flow.

The second model has the advantage of solving another difficulty: to keep $\Sigma < \Sigma_{\max}$ for the duration of low states the disc instability model requires very low accretion rates in quiescence (compare with eq. 4 below): $\dot{M} \lesssim 5 \times 10^{12} - 10^{13}$ g s$^{-1}$ whereas observations of UV and hard X-ray radiation in quiescence suggest $\dot{M} \sim 5 \times 10^{14}$ g s$^{-1}$. A similar problem arises in the case of SXTs. It is worth investigating if the Meyer–Meyer-Hofmeister model could apply also in this case. It is clear that the Livio and Pringle (1992) model cannot work in systems containing black-holes.

Finally, both models may suppress "inside-out" outbursts by removing the "inside".

## 3. The thermal-tidal versus the enhanced mass-transfer instability models

The SU UMa systems are characterised by the appearence of so-called "superoutbursts": stronger outbursts that are longer than "normal" ones. Typically they last more than two weeks. During these events photometric disturbances ("superhumps") are observed in the optical light-curve with periods a few percent longer than the orbital one. Studies by Whitehurst (1988), Whitehurst and King (1991), Hirose and Osaki (1990) and Lubow (1991a,b) established that the superhump is due to a tidal distortion of the outer disc resulting from the presence of the 3:1 resonance inside the disc. A short but comprehensive review may be found in King (1994). The presence of a 3:1 resonance inside the disc requires mass ratios (secondary/primary) $q = M_2/M_1 < q_{\mathrm{crit}} \approx 0.25 - 0.33$. This explains why all SU UMa systems are found with periods $\lesssim 3$ hr. Observations of "superhumps" in some black-hole SXTs are a nice confirmation of the model, since one expects those systems to have low mass-ratios.

The superhump is therefore well explained, but the origin of superoutbursts is subject to debate (e.g. Ichikawa, Hirose and Osaki 1993; Whitehurst 1994).

According to Osaki (1989) the superoutburst is a pure disc phenomenon. In his model, normal outbursts occuring between superoutbursts (the "supercycle") are due to the thermal disc instability. Since, as I have mentioned above, normal outbursts are not efficient in causing accretion on to the central body, both the disc mass and radius grow during the supercycle. When the radius is large enough to contain the 3:1 resonance, the disc becomes ec-centric due to a tidal instability. The tidal instability then enhances removal of angular momentum from the outer part of the disc allowing accretion of a large fraction of matter accumulated during the supercycle. This last point, as Whitehurst (1994) pointed out, assumes a very efficient coupling ("viscosity") between the outer and inner disc.

The tidal-thermal instability model makes several, very definite, predictions which can be tested by observation. First, it predicts a secular growth of disc radius (in a 'normal' cycle the radius first increases then decreases). Second, it predicts a rapid decrease of the superhump period during the superoutburst (Whitehurst 1994). Third, it requires the length of the 'normal' cycle to increase during the supercycle. Observations of the disc radius of the system Z Cha show behaviour opposite to the predictions of the tidal-thermal instability model (Smak 1991). There is not enough data to test the second prediction. Finally, observation of VW Hyi shows that generally the cycle length increases during the supercycle but decreases in the final cycles just before the superoutburst. The data in both cases show a large scatter but no tendency to follow the predictions of the Osaki model can be detected. One may conclude, that this model is not confirmed by observations.

On the other hand, as Smak (1991) has pointed out, a slight enhancement of the mass transfer during the last outbursts is capable of making the cycle length decrease. Such an enhancement is indeed observed (Vogt 1983), and as Smak (1991) has stressed, it is seen only in outbursts preceding the superoutbursts – those with decreasing length. van der Woerd nad van Paradijs (1987) concluded that the recurrence behaviour of VW Hyi suggests that superoutbursts are due to enhanced mass transfer from the secondary.

The (moderate) mass-transfer enhancement model for superoutbursts was proposed by Whitehurst (1994). Here the enhanced mass transfer explains the length of the superoutburst. The exact cause of this enhancement is not known, but it is natural to think that it is due to the effects of illumination by the light emitted during normal outbursts. This model has the advantage of being supported by observation. In addition, the mass transfer enhancement episodes are independent of $q$, so that only for $q \lesssim 0.33$ would the resulting long outburst develop a superhump. U Gem showed a very long (45 day) outburst which was not classified as a superoutburst because a superhump was absent. The mass ratio for this system is $\sim 0.46$. This type of dwarf nova behaviour cannot be explained by the Osaki model. In any case short-time mass-transfer variations in dwarf-novae are not well studied (see Wood *et al.* 1994).

3.1. WZ SGE AND RELATED SYSTEMS

WZ Sge has the particularity of showing only superoutbursts with a recurrence time of $\sim 33$ years. A few other systems show similar but less extreme behaviour. The properties of WZ Sge are most probably connected to its short period: it is very close to the minimum period of CVs distribution. The mass transfer in this system, as estimated by Smak (1993) is low $\sim 2 \times 10^{15}$ g s$^{-1}$ and is close to the one obtained from evolution models (Kolb 1994, private communication). It is no more than 2.5 times less than that of e.g. Z Cha (Kolb 1994). The disc instability model requires WZ Sge to have an extremely low quiescent viscosity: $\alpha_{\rm cold} < 5 \times 10^{-5}$ (Smak 1993) as compared to the usual value of $\sim 0.01$. The physical reason for such a small value is not known, so the explanation of the WZ Sge phenomenon (Osaki 1994) by a low viscosity is obviously not satisactory. The disc instability model for SXTs also requires very low values of viscosity (see below).

## 4. Models for transient outbursts of low-mass X-ray binaries

4.1. DISC INSTABILITY MODEL FOR SOFT X-RAY TRANSIENTS

The dwarf-novae–SXTs connection was analysed by van Paradijs and Verbunt (1984). Cannizzo, Wheeler and Ghosh (1985) and Lin and Taam (1984) proposed that the disc instability model used to describe dwarf nova outbursts may explain the SXT phenomenon.

Comparing the "normal" dwarf novae with SXTs one realises immediately that they share only one property: the rise–time to outburst is similar in the two classes of events. Other characteristics are totally different: decay times as well as recurrence times are much longer in SXTs. Light curves of SXTs are rather similar to superoutbursts of WZ Sge. Indeed both types of objects shared the erroneous name of *Nova* because of the shape of their light-curves.

One should therefore expect the disc instability model for SXTs to encounter difficulties similar to those appearing in the case of SU UMa's and WZ Sge.

4.1.1. *The Mineshige and Wheeler model*
Up to now there have been only two detailed calculations modelling SXTs by the disc-instability model: Huang and Wheeler (1989) and Mineshige and Wheeler (1989). In a recent paper (Cannizzo 1994) a new study by Cannizzo, Chen and Livio was announced as being in preparation.

I will discuss here only the Mineshige and Wheeler paper (1989; hereafter MW) because the inner disc boundary in Huang and Wheeler (1989) is assumed to be at $10^8$ cm which is certainly at least one order of magnitude too large for a "realistic" model of a SXT. However, I have to comment on a claim by Chen, Livio and Gehrels (1993) and Mineshige *et al.* (1994) that the observed rise of hard X-rays preceding the rise in soft X-rays (Ricketts, Pounds and Turner 1975; Lund 1993) in some SXTs is consistent with the "inside-out" character of outbursts predicted by the disc instability model. First, there is no unique model for hard and soft X-ray emission in the inner regions of accretion discs around compact objects, so in itself an "inside-out" character of an outburst proves nothing. Second, whereas it is true that in the case of dwarf novae, models assuming constant values of $\alpha$ in quiescence get "inside-out" oubursts for low accretion rates (Smak 1984), the MW model for SXTs succeeds in igniting the inner disc regions *only* for "outside-in" outbursts. This is clearly explained in MW and will be discussed below. This point is nevertheless extremely confusing because of a sentence in Mineshige *et al.* (1994) which suggests that Fig. 7 in MW represents a "thermal instability initiated from the inner part", while, in discussing the same figure, MW ask the reader to note that "this is an 'outside-in' burst...". In this particular model (for which the recurrence time is about 7 months, and the peak bolometric luminosity is $< 10^{35}$ erg s$^{-1}$ – parameters not very close to those of SXTs) the inner disc never falls to the 'cold' state because the cooling wave is reflected at $r \approx 20 r_{\rm in}$ as a outward propagating heating wave. As explained in MW, before this reflected heating wave manages to propagate to more than $r \approx 20 r_{\rm in}$ the new "outside-in" heating front created at $r \approx 800 r_{\rm in}$ propagates inward and merges with the hot inner region. The main outburst is therefore of the "outside-in" type.

On the other hand Huang and Wheeler (1989) get "inside-out" outbursts but their "inside" is at $10^8$ cm, obviously not relevant to the problem of the soft/hard X-ray priority.

For a constant $\alpha$ the MW $T_{\rm eff} - \Sigma$ relation at $10^7$ cm shows no unstable branch below $\sim 3 \times 10^4$ K. To get an "S-curve" they have to tune up the viscosity prescription. MW use the $\alpha = \alpha_0 \left(H/r\right)^n$ viscosity prescription and try to adapt both $\alpha_0$ and $n$ to obtain light curves corresponding to SXTs. They never succeed in getting recurrence time longer than $\sim 16$ years. The sequences of outbursts show only a vague resemblance to observations (see their figs. 5, 6, 8 and 9). MW use 21 grid points in their calculation. It is therefore interesting to compare their results with those of Cannizzo (1993b) who shows the difference between calculations using 25 and 100 grid points. One has the impression that MW results suffer seriously from too coarse a resolution.

The best results, in the sense of amplitudes and recurrence times closest to observations, are obtained for $n > 1$ and $\alpha_0 \geq 10$. With the prescription used by MW, the viscous time $\tau_{\rm vis} \approx (H/r)^{-2}(1/\alpha\Omega)$ (see e.g. Frank, King and Raine 1992) is proportional to $r^{(1-n)/2}$. For $n > 1$, $\tau_{\rm vis}$ is *decreasing* with radius so that even for low accretion rates (long accumulation times) MW get only "outside-in" outbursts.

As for WZ Sge the $\alpha$ required by the disc instability model is very small in quiescence. MW do not give its value at the disc inner edge ($3.16 \times 10^6$ cm) but at $10^7$ cm it is $8 \times 10^{-5}$ for $\alpha_0 = 3.16 \times 10^3$, $n = 2$ and $2 \times 10^{-4}$ for $\alpha_0 = 10^2$, $n = 1.5$.

From Fig. 10 in MW one sees that the accretion rate at the inner disc radius in quiescence is extremely low $\lesssim 10^6$ g s$^{-1}$. One can show that this is a necessary consequence of the disc instability model. Indeed, this model requires that the surface density remain everywhere below the critical surface density $\Sigma_{\rm max}$ during the quiescence. For black-hole SXTs the recurrence time is longer than 10 years. This implies that viscosity at the inner edge of the disc satisfies the inequality:

$$\nu \lesssim 10^6 t_8^{-1} r_7^2 \quad {\rm cm}^2 \ {\rm s}^{-1}. \tag{2}$$

If one uses the formula for $\Sigma_{\rm max}$ from Smak (1992):

$$\Sigma_{\rm max} = 16.22 r_{10}^{1.11} m_1^{-0.37} \alpha^{-0.79} \quad {\rm g \ cm}^{-2}, \tag{3}$$

one gets an inequality for the value of the accretion rate at the inner boundary of the disc:

$$\dot{M}_{\rm in} \lesssim \frac{8\pi}{3} \nu \Sigma_{\rm max} \approx 2.76 \times 10^4 t_8^{-1} r_7^{3.11} m_{10}^{-0.37} \alpha^{-0.79} \quad {\rm g \ s}^{-1} \tag{4}$$

in very good agreement with MW. Notice that I used a constant $\alpha$ but this does not affect the result.

Several black-hole SXTs have been observed by GINGA (Mineshige *et al.* 1992) and ROSAT (see the article by Verbunt in these proceedings and Verbunt *et al.* 1994). Two systems, A0620-00 and V404 Cyg, have been detected at levels corresponding to accretion rates of at least $\sim 1.4 \times 10^{10}$ g s$^{-1}$ for A0620-00 and of $3 \times 10^{12}$ g s$^{-1}$ for V404 Cyg. Those accretion rates are several orders of magnitude larger than the values predicted by the disc instability model of MW. Detections of X-ray radiation from quiescent SXTs contradict the disc instability model, at least in the version proposed by MW (Mineshige *et al.* 1992).

Other observations by GINGA and ROSAT which provide upper limits on X-ray radiation from quiescent SXTs contradict the mass-transfer instability model proposed by Hameury, King and Lasota (1986), as I discuss later.

One of the main ingredients of the disc instability model is the presence of an inward propagating "cooling wave" which brings the hot disc back to the (non-equilibrium) cool state through series of quasi-equilibria. As noticed by Cheng *et al.* (1992) the *HST/FOS* spectra of the so called X-ray Nova Muscae 1991 show no evidence for the propagation of a cooling front in the accretion disc of this system. Chen *et al.* (1993) suggest that this is due to hard X-ray heating of the outer disc regions. Effects of the disc illumination by X-ray on SXTs outbursts still await a detailed treatment (see Mineshige, Kim and Wheeler 1990; Kim, Wheeler and Mineshige 1994; Mineshige *et al.* 1994).

The physics of the hot inner disc around black holes and neutron stars is still poorly known. For recent results on this subject see Abramowicz *et al.* (1994), Chen and Taam (1994), Milsom, Chen and Taam (1994).

4.2. THE MASS–TRANSFER INSTABILITY MODEL FOR SXTS

A mass-transfer instability model for SXTs was proposed in a series of papers by Hameury, King and Lasota (1986, 1987, 1988, 1990; hereafter HKL). HKL realised that the mass transfer from an X-ray heated secondary star is unstable for a range of accretion rates between two critical values. The lower value corresponds to an illuminating X-ray flux comparable to the intrinsic stellar flux and is typically $\sim 10^{12}$ - $10^{14}$ g s$^{-1}$ depending on the binary parameters. The upper critical accretion rate above which X-ray heating dominates is $\sim 10^{16.5}$ g s$^{-1}$ for $P_{\rm orb} \sim 3$ h and increases more or less linearly with the orbital period. The two accretion rate limits, expressed in terms of X-ray luminosities, correspond closely to the limits of the "luminosity gap" of steady LMXBs (White, Kaluzienski and Swank 1984, Johnston - these proceedings) which is a very attractive feature of the model.

According to HKL, during the quiescence accretion rates are slightly smaller than the lower critical value. It is assumed that a substantial amount of the accretion luminosity is emitted in hard ($E > 7$ keV) X-rays which are able to increase the companion's effective temperature. The secondary expands under the effect of illumination and at some moment enters the unstable range of accretion rates. A mass-transfer runaway produces the SXT outburst. When the accretion disc thickness is large enough to shield the region around the Lagrange $L_1$–point, the runaway stops. The secondary contracts and the system goes back to quiescence. During the decay from outburst the $L_1$ region may become uncovered and produce new outbursts.

This version of the mass-transfer instability model suffers from one major difficulty: it requires the secondary star to react to the effects of X-ray illumination on time-scales of the order of months. The secondary's heated layer must be able to expand by at least a fraction of the atmospheric scale-height in less $10^6 - 10^7$ s. For example, GINGA observations of GS 2000+25 150 days *before* the outburst have not detected X-rays between 1.2 and 37.1 keV implying an upper limit $\sim 6 \times 10^{33}$ erg s$^{-1}$ (Mineshige *et al.* 1992). This value is not in contradiction with the HKL model, but let us assume that the actual X-ray luminosity is much lower. In itself this would not invalidate the model: one could imagine that hard X-rays appear only three months before the outburst, due for example to an instability in the *inner* disc. The problem is whether the outer layers of the star may expand by a scale-height during three months.

Gontikakis and Hameury (1993) showed (see also Ritter in these proceedings) that for a main sequence star with a mass less than $\sim 1 M_\odot$ the expansion time is always longer than $\sim 10^2$ years. For a $0.8 M_\odot$ main-sequence secondary the characteristic time is $\sim 10^4$ years. Low-mass main-sequence stars possess deep, massive convective zones which react *globally* to surface heating on their thermal time-scale. The local thermal time-scale introduced by HKL (1986) makes no sense in a convective envelope. Only massive main-sequence stars and evolved low-mass stars (subgiants or stripped giants) can react to X-ray illumination on times-scales as short as $10^6$ s.

The orbital period of $\sim 5.1$ hr determined for J0422+32 seems to rule out the mass-transfer instability as a cause of its ouburst since the companion star, if at main sequence, should have a mass of $\sim 0.56 M_\odot$.

One could of course try to defend the model by saying that SXTs have the particularity of containing evolved stars at short orbital period or to say that the HKL model applies only to systems with $P_{\rm orb} \gtrsim 9$ hr but unless there is an independent reason to believe that this is true, the HKL mass-transfer instability model cannot describe the SXT phenomenon.

Chen, Livio and Gehrels (1993) and Augusteijn, Kuulkers and Shaham (1993) have proposed models which combine elements of both type of models. Although the proposed versions were not confirmed by observations (Callanan *et al.* 1994; Chevalier and Ilovaisky 1994), this is probably a direction to follow.

I am grateful to Wolfgang Duschl, Jean-Marie Hameury, Andrew King, Uli Kolb, Darragh O'Donoghue, Hans Ritter and Joe Smak for useful discussions, advice and comments. I thank Mike Garcia and Janet Wood for sending me their work prior to publication.

**References**

Abramowicz, M.A. *et al.* (1995) *Astrophys. J. Lett.*, 1 January
Augusteijn, T., Kuulkers, E. and Shaham, J. (1993) *Astron. Astrophys.* **279**, L13
Callanan, P.J. *et al.* (1994) *Astrophys. J.*, submitted

Cannizzo, J.K. (1993a) in *Accretion Disks in Compact Stellar Systems*, ed. J. Craig Wheeler, World Scientific Publishing, Singapore, p.6
Cannizzo, J.K. (1993b) *Astrophys. J.* **419**, 318
Cannizzo, J.K. (1994) *Astrophys. J.* **435**, 389
Chen, W., Livio, M. and Gehrels, N. (1993) *Astrophys. J.* **408**, L5
Chen, X. and Taam, R.E. (1994) *Astrophys. J.* **431**, 732
Cheng, F.H. et al. (1992) *Astrophys. J.* **397**, 664
Chevalier, C. and Ilovaisky, S. (1994) *preprint OHP*
Frank, J., King, A.R. and Raine, D.J. (1992) *Accretion Power in Astrophysics*, 2nd Edition, CUP, Cambridge
Gontikakis, C. and Hameury, J.M. (1993) *Astron. Astrophys.* **271**, 118
Hameury, J.M., King, A.R. and Lasota, J.P. (1986) *Astron. Astrophys.* **162**, 71
Hameury, J.M., King, A.R. and Lasota, J.P. (1987) *Astron. Astrophys.* **171**, 140
Hameury, J.M., King, A.R. and Lasota, J.P. (1988) *Astron. Astrophys.* **192**, 187
Hameury, J.M., King, A.R. and Lasota, J.P. (1990) *Astrophys. J.* **353**, 585
Hirose, M. and Osaki, Y. (1990) *Publ. Astro. Soc. Japan* **42**, 135
Huang, M. and Wheeler, J.C. (1989) *Astrophys. J.* **343**, 229
Ichikawa, S., Hirose, M. and Osaki, Y. (1993) *Publ. Astro. Soc. Japan* **45**, 243
Kim, S.W., Mineshige, S. and Wheeler, J.C. (1994) in *The Evolution of X-ray Binaries*, eds. S.S. Holt and C.Day, AIP Conference Proceedings, New York, in press
King, A.R. (1994) in *The Evolution of X-ray Binaries*, eds. S.S. Holt and C.Day, AIP Conference Proceedings, New York, in press
Lin, D.N.C. and Taam, R.E (1984) in *High Energy Transients in Astrophysics*, ed. S.E. Woosley, AIP Conf. Proc. No. 115, p. 83
Livio, M. and Pringle, J.E. (1992) *Mon. Not. R. astr. Soc.* **259**, 23P
Lubow, S.H. (1991a) *Astrophys. J.* **381**, 359
Lubow, S.H. (1991b) *Astrophys. J.* **381**, 368
Ludwig, K., Meyer-Hofmeister, E. and Ritter, H. (1994) *Astron. Astrophys.* **290**, 473
Lund, N. (1993) *Astrophys. J. Sup. Ser.* **97**, 289
Meyer, F. and Meyer-Hofmeister, E. (1994) *Astron. Astrophys.* **288**, 175
Milsom, J.A., Chen, X. and Taam, R.E. (1994) *Astrophys. J.* **421**, 668
Mineshige, S. and Osaki, Y. (1983) *Publ. Astro. Soc. Japan* **35**, 377
Mineshige, S. and Wheeler, J.C. (1989) *Astrophys. J.* **343**, 241
Mineshige, S., Kim, S.W. and Wheeler, J.C. (1990) *Astrophys. J.* **358**, L5
Mineshige, S., Yamasaki, T. and Ishizaka C. (1993) *Publ. Astro. Soc. Japan* **45**, 707
Mineshige, S. et al. (1994) *Astrophys. J.* **426**, 308
Mineshige, S. et al. (1992) *Publ. Astro. Soc. Japan* **44**, 117
Osaki, Y. (1989) *Publ. Astro. Soc. Japan* **41**, 1005
Osaki, Y. (1994) in *Theory of Accretion Disks – 2*, eds. W. Duschl, J. Frank, E. Meyer-Hofmeister, F. Meyer and W. Tcharnuter, Kluwer, Dordrecht, p. 93
van Paradijs J. and Verbunt, F. (1994) in *High Energy Transients in Astrophysics*, ed. S.E. Woosley, AIP Conf. Proc. No. 115, p. 49
Ricketts, M.J., Pounds, K.A. and Turner, M.J. (1975) *Nature* **257**, 657
Shakura, N.I. and Sunyaev R.A. (1973) *Astron. Astrophys.* **24**, 337
Smak, J.I. (1984) *Acta Astron.* **34**, 161
Smak, J.I. (1991) *Acta Astron.* **41**, 269
Smak, J.I. (1992) *Acta Astron.* **42**, 323
Smak, J.I. (1993) *Acta Astron.* **43**, 101
Verbunt, F.et al. (1994) *Astron. Astrophys.* **285**, 903
Vogt, N. (1983) *Astron. Astrophys.* **118**, 95
White, N., Kaluzienski, J.L. and Swank, J.H. (1984) in *High Energy Transients in Astrophysics*, ed. S.E. Woosley, AIP Conf. Proc. No. 115, p. 31
Whitehurst, R. (1988) *Mon. Not. R. astr. Soc.* **232**, 35
Whitehurst, R. (1994) *Mon. Not. R. astr. Soc.* **266**, 35
Whitehurst, R. and King, A.R. (1991) *Mon. Not. R. astr. Soc.* **249**, 25

van der Woerd, H. and van Paradijs, J. (1987) *Mon. Not. R. astr. Soc.* **224**, 271
Wood, J. *et al.* (1994) *preprint*